%% file: main.tex
\documentclass[final,twocolumn,prl,aps,floats,amsfonts,amssymb,stmaryrd,showpacs]{revtex4-2}

\usepackage{graphicx}
\usepackage{dcolumn}
\usepackage{mathtools}
\usepackage{bm}
\usepackage{hyperref}
\usepackage{xcolor}
\usepackage[caption=false]{subfig}
\usepackage{subcaption}
\usepackage{natbib}
\usepackage{cleveref}
\usepackage{rotating}
\usepackage{adjustbox}
\usepackage{tikz}
\usetikzlibrary{shapes}
\usepackage{amsmath}
\usepackage{booktabs}

\usepackage{etoolbox}
\usetikzlibrary{plotmarks}
\usepackage{MnSymbol}

\usepackage{comment}

\definecolor{darkgreen}{RGB}{0.0, 100, 0.0}
\definecolor{purple}{RGB}{128, 0.0, 128}
\definecolor{orange}{RGB}{255, 149, 0.0}
\definecolor{grey}{RGB}{127, 127, 127}

\definecolor{blueplot}{RGB}{12.0, 93, 165}
\definecolor{greenplot}{RGB}{0.0, 185, 69}
\definecolor{orangeplot}{RGB}{255, 149, 0.0}
\definecolor{brownplot}{RGB}{140, 86, 75}
\definecolor{pinkplot}{RGB}{227, 119, 194}

\usetikzlibrary{shapes.geometric}

\begin{document}

\title{Universal self-similar evolution of two-dimensional acoustic turbulence in Bose-Einstein condensates.}

\author{Guillaume Costa}
\email{guillaume.costa@oca.eu}
\author{Sergey Nazarenko}
\author{Giorgio Krstulovic}

\affiliation{Universit\'{e} C\^{o}te d'Azur, CNRS, Institut de Physique de Nice (INPHYNI), 17 rue Julien Lauprêtre, 06200 Nice, France}

\date{\today}

\begin{abstract}
    When driven out of equilibrium, a Bose-Einstein condensate develops nonlinearly interacting density waves that trigger a turbulent cascade, transferring energy toward small scales.
    In this Letter, we investigate the nonstationary evolution of solutions to the two-dimensional Gross–Pitaevskii equation (GPE). Through numerical simulations of both the GPE and the corresponding Wave Kinetic Equation (WKE), we identify self-similar solutions relevant to turbulence in atomic and polariton Bose–Einstein Condensates. These solutions correspond to a new type of non-thermal fixed point and exhibit characteristics of both first and second kind self-similarity. In particular, we show that the dynamics of the propagating front is universal, governed by a dimensionless universal constant $\beta$, which we determine numerically.
\end{abstract}

\maketitle

Self-similar evolution, in which the dynamical variables of a system evolve obeying some scaling relation, is a common phenomenon in nature where complex physics is encoded in somehow simple laws. Perhaps, one the most simple and famous examples in physics is the diffusion of an ink stain, where the size of the stain scales as the square root of the time. Another notorious example is the spherical shock expansion after a nuclear explosion. In early era of nuclear test, G.~I.~Taylor was able to give a correct estimate of the energy of the explosion by applying  dimensional analysis and the self-similar assumption~\cite{taylor1950formation1, taylor1950formation2}. Perhaps more intricate is the implosion of a sphere of vacuum after a sudden removal of internal fluid. In this case, the energy outside the sphere is in practice infinite and whereas the process is still self-similar, scaling exponents cannot be found easily. The two previous examples lie in two different classes of self-similarity. 
According to the Zel{'}dovich classification~\cite{zel2002physics}, if the evolution of a system can be fully characterized using dimensional arguments and conservation laws, it is said to exhibit self-similarity of the first kind. In contrast, for self-similarity of the second kind, universal scaling exponents are typically determined by a nonlinear eigenvalue problem, whose solution is usually obtained numerically.

In most nonlinear, non-equilibrium evolutions, self-similarity emerges as a consequence of the scale invariance of the governing equations. The general question is: how does an initially narrowly supported condition—whether in physical or Fourier space—propagate in time?  
Dimensional analysis and the use of conservation laws often combine together with self-similar hypothesis to develop a simple, but yet powerful prediction of such complex non-steady physical systems. In a very general manner, when the dynamics is studied in Fourier space, the complete evolution of such systems is characterized by the evolution of a front towards the ultra-violet (UV) or the infrared (IR), and by the existence of universal self-similar functions. Such self-similar functions often display a power-law behavior in one end and sharp decay in the end where the front is. In such regimes, details of the initial conditions are  irrelevant, and this behavior is indicative of the presence of \textit{non-equilibrium attractors}, often referred to as \textit{non-thermal fixed points}~\cite{schmied2019non} in the context of field theory.
Famous examples are finite-time UV singularities of some fluid-like models~\cite{connaughton2004warm, grebenev2014self, thalabard2015anomalous, bos2012developing, costa2025behind, campolina2018chaotic, pikeroen2024tracking} and the IR blow-up of the wave kinetic equation (WKE) describing the dynamical condensation process of Bose-Einstein condensates (BEC)~\cite{Semikoz1995,zhu2023self}.
    \begin{figure*}[!htb]
            \centering
            \includegraphics[width = \linewidth]{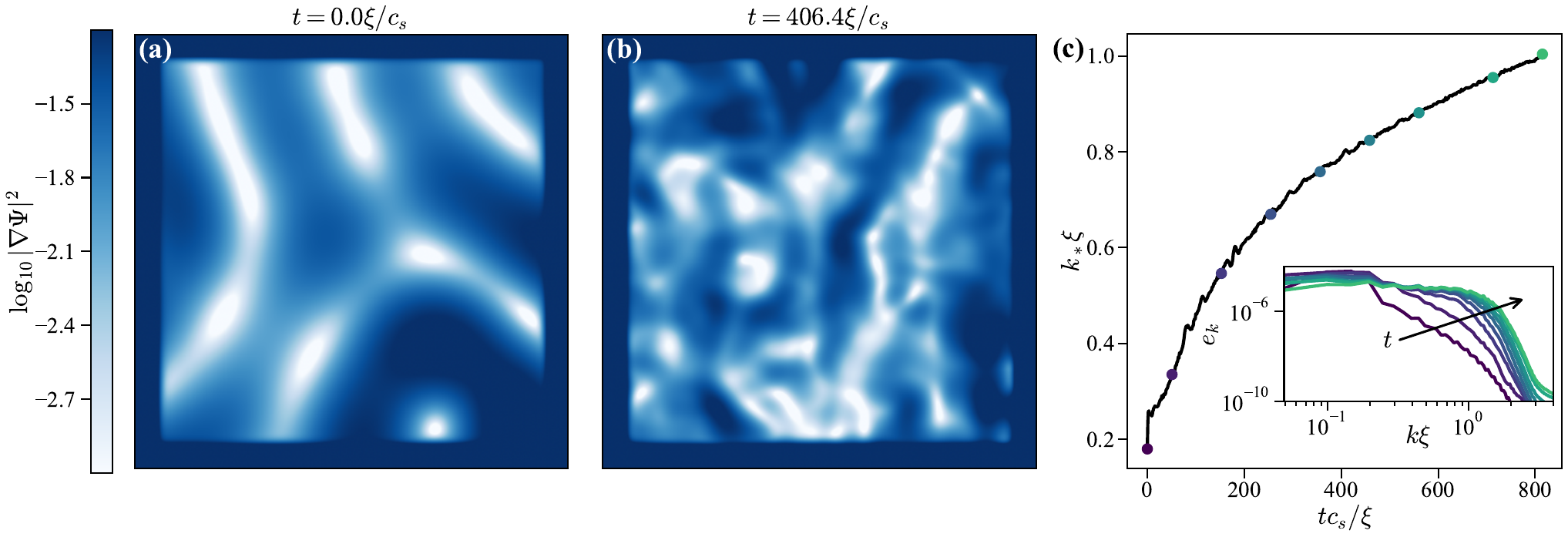}
            \caption{\textbf{(a)} \& \textbf{(b)} Snapshots of GPE simulations with trap confinement of size $75\xi$ highlighting the formation of finer and finer structures during the time evolution. \textbf{(c)} \textbf{Main panel:} Time evolution of $k_*$ supporting the formation of smaller and smaller scale structures. \textbf{Inset:} Time evolution of the energy spectrum for the GPE  with trap, highlighting the propagation of a front in Fourier space.}
            \label{fig:snapshots}
    \end{figure*}
In general terms, WKEs describe the temporal evolution of the energy spectrum of dispersive wave equations in the weakly nonlinear (and infinite box size) limit. They are derived in the framework of the weak wave turbulence theory (WWTT). Despite their apparent complicated mathematical structure (see later), 
they often admit simple power-law solutions. In the absence of forcing and dissipation, thermal solutions known as the Rayleigh-Jeans (RJ) spectrum are readily found using the invariants of the WKE. More interesting, in non-equilibrium settings, and when forcing and dissipation scales are well separated, WKE admits power-law constant flux solutions known as Kolmogorov-Zakharov (KZ) solutions~\cite{zakharov1965weak, nazarenko2011wave}. As we will discuss later, the physical properties of KZ solutions play a crucial role in determining the kind of self-similarity. Over the last years, the WWTT has been proven as a powerful theoretical and numerical tool, and it has been applied to systems as diverse as gravitational waves~\cite{galtier2017turbulence}, elastic plates~\cite{during2006weak}, magnetohydrodynamics~\cite{galtier2000weak}, internal gravity waves in the ocean~\cite{caillol2000kinetic, galtier2003weak}, and BECs~\cite{zhu2023direct}.

In the case of atomic BECs, there has been a tremendous experimental progress over the last ten years in realizing 2D~\cite{galka2022emergence} and 3D~\cite{navon2016emergence} homogeneous forced and dissipated wave turbulence settings, studying non-equilibrium steady state spectra, as well as non-thermal fixed points (self-similar evolution). Note, however, that whereas recent 2D atomic BEC experiments have realized wave turbulent settings to study the inverse particle cascade and the acoustic direct energy cascade, polariton BECs have mainly focused on vortex dynamics and hydrodynamic turbulence~\cite{lagoudakis2008quantized, panico2023onset, byrnes2014exciton}. One important difficulty is the polariton finite life time~\cite{byrnes2014exciton, carusotto2013quantum}, which leads to important losses and therefore to strong dissipative effects.

In this Letter, we study the self-similar evolution of an initially large-scale 2D BEC in the acoustic regime where a strong condensate is present. To model losses in polaritons, we  also consider a linear damping term and show that, regardless of this dissipative effect, self-similarity is preserved and the universal self-similar form remains the same. We provide exact analytical predictions within the WWTT and confirm our results numerically.

We start by considering the 2D Gross-Pitaevskii equation (GPE), which describes the dynamics of atomic and polariton BECs. The GPE expressed in terms of the speed of sound $c_s$ and the healing length $\xi$ reads:
    \begin{equation}\label{Eq:GP}
    i \frac{\partial \Psi}{\partial t} = \frac{c_s}{\sqrt{2} \xi}\left[-\xi^2 \nabla^2 + \frac{|\Psi|^2}{\rho_0} - 1\right] \Psi - V_{\rm trap}({\bf x})\Psi- i \gamma\Psi.
    \end{equation}
where $\rho_0$ is the bulk density, $V_{\rm trap}$ is the external potential confining the BEC, and $\gamma$ is the damping coefficient. Note that for polaritons, typical experimental values of the dimensionless damping are $\gamma \xi/c \sim0.01 -0.1$~\cite{panico2023onset} and for atomic BEC is zero. 

When Eq.\eqref{Eq:GP} is linearized around a constant density $\rho_0$, waves propagate with the Bogoliubov dispersion relation
\begin{equation*}
    \omega_k = c_s k \sqrt{1 + \left(k\xi\right)^2/2}\approx c_s k \left(1+ \frac{\xi^2 k^2}{4}\right)\quad {\rm for\,} k\xi\ll1 ,
\end{equation*}
which at large scales propagate as acoustic modes.

\begin{figure}[!ht]
    \centering
    \includegraphics[width=1.\columnwidth]{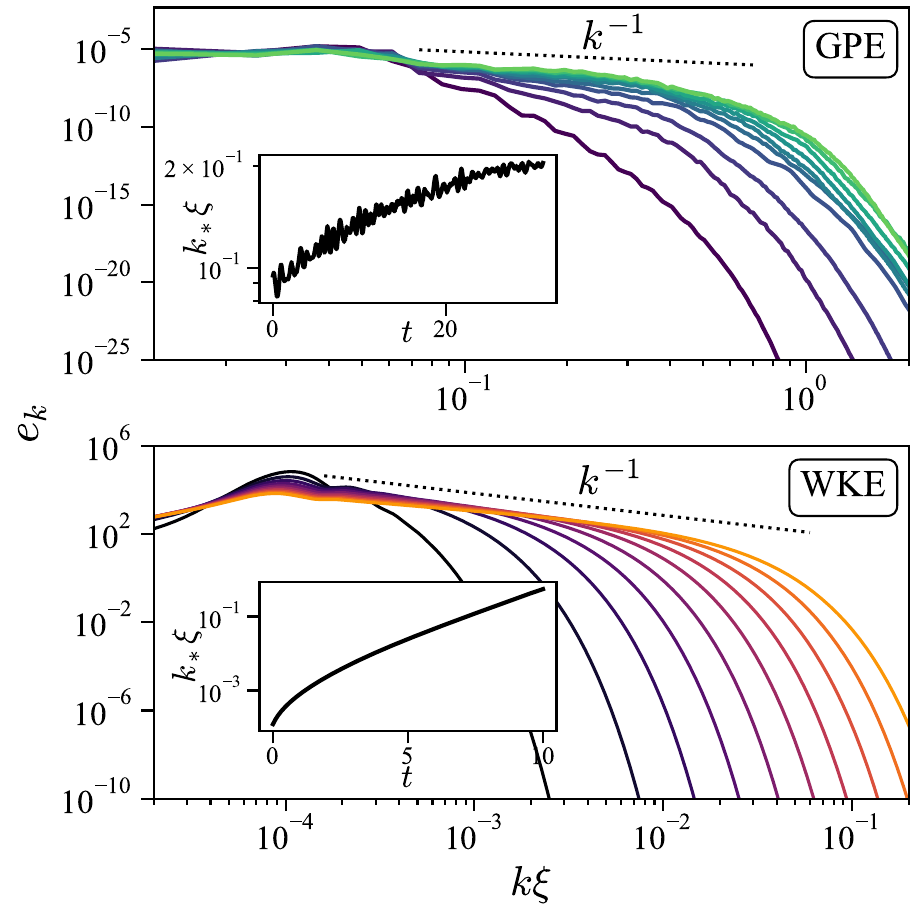}
    \caption{Snapshots of the energy spectrum for the GPE (top) and WKE (bottom) highlighting the propagation of a front leaving a KZ spectrum $e_k \propto k^{-1}$ in its wake. The insets show their respective $k_*(t)$ from Eq.~\eqref{eq:defkstar}}
    \label{fig:GP_WKE_spec}
\end{figure}

We first consider the GPE using a homogeneous square trap of size $75\xi$, which is comparable to the size of current experiments~\cite{panico2023onset}, and use the pseudospectral code FROST~\cite{KrstulovicHDR} to integrate the equations (see Supplemental Information (SI) for more details).
We first obtain the ground state by imaginary time evolution and then perturb it superimposing weak random Bogoliubov wave at large scales, as displayed in Fig.~\ref{fig:snapshots}a showing the kinetic energy density.  As time increases, the system develops progressively finer structures, clearly visible in Fig.~\ref{fig:snapshots}b. A simple way of quantifying the typical scale of such structures is to use the energy-weighted average 
\begin{equation}
    k_*(t) = \left({\int_0^\infty k^3e_k\mathrm{d}k}\Big/{\int_0^\infty k e_k\mathrm{d}k} \right)^{1/2},
    \label{eq:defkstar}
\end{equation}
where $e_k=2c^2\xi^2\int_{|{\bf k}|=k}k^2 |\hat{\psi}_{\bf k}|^2\mathrm{d}{\bf k}$ is the energy spectrum and $\hat{\Psi}_{\bf k}$ the Fourier transform of $\Psi$ (see SM for details). The choice of powers in Eq.~\eqref{eq:defkstar} will become clearer later. Note that $k_*$ can be interpreted as the wavevector at which energy is concentrated i.e. the inverse integral scale. The time evolution of $k_*$, averaged over 140 realizations of the initial condition, is reported in Fig.~\ref{fig:snapshots}c, which shows an increasing behavior compatible with the previously observed formation of small scale structures. Moreover, the energy spectrum exhibits a front propagating from small to large wavectors, as shown in the inset of Fig.~\ref{fig:snapshots}c.
The small size of the trap in the latter simulation makes it difficult to uncover universal features of a self-similar expansion. We therefore perform a simulation without trap in a periodic box of size $1024\xi$. The temporal evolution of the energy spectrum is displayed in Fig.~\ref{fig:GP_WKE_spec} (top panel), confirming dynamics consistent with self-similar solutions. Small scales are developed through a propagating front, leaving in its wake a power-law range $e_k \sim k^{-\alpha}$, with $\alpha \approx 1$.

A deeper theoretical understanding of the spectrum evolution can be obtained using the WWTT developed for large-scale Bogoliubov waves in two dimensions~\cite{zakharov1970spectrum, costa2025stability, griffin2022energy}. In this framework, the evolution of the energy spectrum is governed by the wave kinetic equation (WKE):
     \begin{equation}
        \begin{aligned}
            \partial_t e_k &= \dfrac{4V_0^2}{\sqrt{6} c_s^2\xi}S_t[e_k] - \gamma e_k,\\
            S_t[e_k] &= \int_0^\infty \left(k_1k_2\right)^{-1} \left(\mathcal{R}_{1,2}^k - 2\mathcal{R}_{k,2}^1\right) \, dk_1, \\
            \mathcal{R}_{1,2}^k &= H(k_2)\left(k^2e_1 e_2 - k_1^2e_2e_k - k_2^2 e_1e_k\right),
        \end{aligned}
        \label{eq:SimpleWKE}
    \end{equation}
where $k_2 = |k-k_1|$, $H$ is the Heaviside function, and $V_0 = 3 \sqrt{c_s/32}$ is the interaction constant arising in the acoustic limit of the GPE~\cite{griffin2022energy}. 
Before proceeding, let us make several comments on the WKE. First, the derivation of the WKE is rigorous only in the absence of forcing and dissipation. The damping term $\gamma e_k$ has thus been added in an ad-hoc manner. Second, it is well known that the WWTT presents  serious mathematical issues for acoustic waves~\cite{zakharov2012kolmogorov, griffin2022energy, costa2025stability}, notably in 2D. A wave kinetic description can nevertheless be obtained by assuming a finite but small healing length $\xi$. This small dispersion limit introduces an explicit dependence on $\xi$ in the prefactor of the 2D collision integral, leading to a dimensional exponent different from the one of the Zakharov-Sagdeev spectrum observed in 3D \cite{zhu2024turbulence}. Third, the total energy of the system is $E = \int_{0}^\infty e_kdk$, and its evolution is simply given by 
\begin{equation}
E(t)=E_0 e^{-t/ \tau_D},\label{eq:Energy_decay}
\end{equation}
where $\tau_D \equiv \gamma^{-1}$ is the characteristic dissipation time and $E_0$ the initial energy. The energy is therefore conserved by the WKE when $\gamma=0$.
Fourth, in the absence of damping, Eq.~\ref{eq:SimpleWKE} admits two types of steady-state solutions, the RJ thermal equilibrium spectrum $e_k \propto k$, and the non-equilibrium KZ solution $e_k = \dfrac{2^{5/4}}{3^{3/4}\pi}\sqrt{P_0 c_s\xi}\,k^{-1}$, where $P_0$ is the energy flux~\cite{griffin2022energy}. This KZ spectrum corresponds to a direct cascade where energy is transferred from large to small scales through nonlinear wave interactions.

We now perform a numerical simulation of the WKE~\eqref{eq:SimpleWKE} using the code WavKinS~\cite{krstulovic2025wavkins} (see SM for details).
We set a large scale initial condition and integrate the WKE without damping using a spectral range spanning for four decades. The dynamics is similar to that observed for the GPE.  In particular, Fig.~\ref{fig:GP_WKE_spec} (bottom panel) shows a clear propagation of a front towards large wavevectors exhibiting now a manifest $k^{-1}$-scaling in its wake, consistent with the KZ solution. 

In the general context of turbulent cascades, the $2d$ acoustic KZ solution is said to have an \textit{infinite capacity}, as it is not integrable in the ultraviolet (UV), and an infinite amount of energy is required to deploy an infinite inertial range. For \textit{infinite capacity} systems, one usually expects a first-kind self-similarity~\cite{nazarenko2011wave} dynamics, fully determined by dimensional analysis and the conservation law. We will see later that here it is not the case; instead the solution emerges as a new type of non-thermal fixed point that have not yet been reported (see for instance \cite{chantesana2019kinetic} for known scaling relations). We shall also notice that this KZ solution also diverges in the infrared (IR), which is inconsistent with the finite energy $E_0$ of the initial condition. Nevertheless, the evolution proceeds only towards the UV, and it is then natural to introduce a IR-cutoff $k_0$ set by initial condition. We therefore suppose that the total energy is given by
\begin{equation}
    E = \int_{k_0}^{\infty} e_k\,\mathrm{d}k.
    \label{eq:IntE}
\end{equation}
and we might expect a universal behavior for $k\gg k_0$ only.

\begin{figure*}[!htb]
    \centering
    \includegraphics[width = \textwidth]{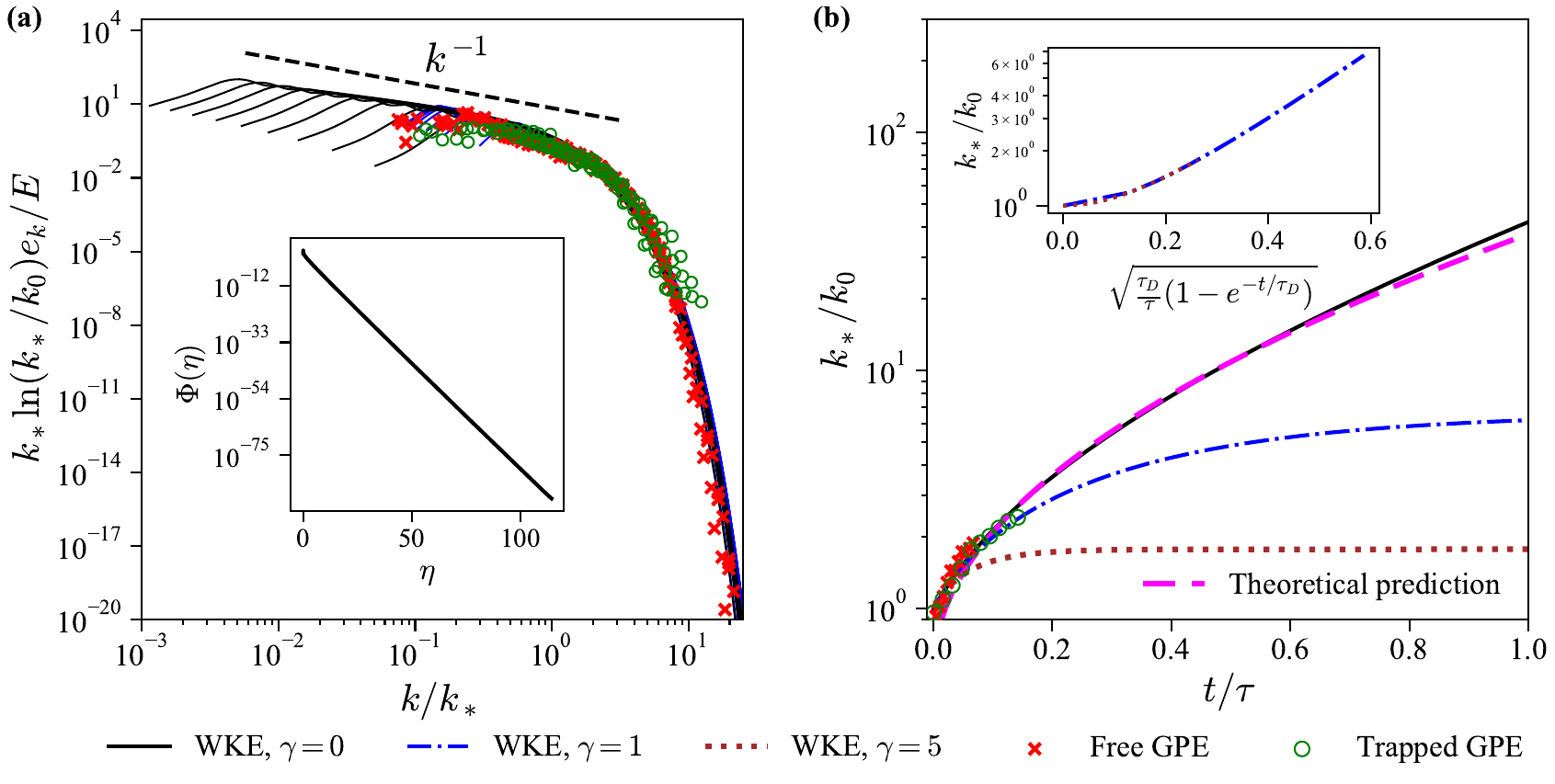}
    \caption{\textbf{(a)} Collapsed self-similar energy spectra. The inset highlights asymptotical behavior of $\Phi$ as $\eta \rightarrow \infty$ confirming the exponential prediction of Eq.~\eqref{eq:Asympt}. \textbf{(b)} Universal behavior of the front propagation confirming the prediction of Eq.~\eqref{eq:kstar}. The magenta dashed line represents the exponential behavior $k_* = k_0\exp\left[\sqrt{\dfrac{2\beta t}{\tau}}\right]$ with $\beta \approx 8.9$. The inset highlights the double exponential regime in presence of dissipation.}
    \label{fig:kstar}
\end{figure*}

We now seek for a self-similar solution of the following form:
\begin{align}
    e_k(t) \equiv E_0f(t) \Phi(\eta)&& \eta = k/k_*(t),
    \label{eq:SelfSimAnsatz}
\end{align}
where $k_*(t)$ represents the position of the front. Using this ansatz in Eq.~\ref{eq:IntE}, the total energy decay law in Eq.~\ref{eq:Energy_decay}, and the expected asymptotic behavior $\Phi(\eta) \underset{\eta \to 0}{\approx} \eta^{-1}$ (to be checked \emph{a posteriori}), simply leads to $f(t) \underset{k_*\gg k_0}{\approx} \exp\left[-t/\tau_D \right]/\left(k_*(t) \ln{\left(\dfrac{k_*(t)}{k_0}\right)}\right)$, determining the amplitude of the spectrum. In addition to the natural decaying term due to dissipation, we notice a non-homogeneous dependence on $k_*(t)$, which is unusual in self-similar evolutions. Note that $k_*$ exactly corresponds, up to a constant, to definition~\eqref{eq:defkstar} and the power of $k$ used in the definition are the lowest ones such that the average does not contain the logarithmic term (see SM).

Substituting the above expression into the WKE~\eqref{eq:SimpleWKE}, and requiring the time dependence to cancel, we get a front evolution equation for $k_*(t) \gg k_0$,
\begin{equation}
    \dfrac{1}{k_*}\dfrac{dk_*}{dt}\ln{\left(\dfrac{k_*}{k_0}\right)} = \dfrac{\beta}{\tau}\exp{\left[-t/\tau_D\right]},
    \label{eq:fronteq}
\end{equation}
together with an equation for the universal self-similar function $\Phi$
\begin{equation}
    -\beta\left[\Phi(\eta) + \eta \Phi'(\eta)\right] = S_t[\Phi]. \label{eq:Selfsimeq}
\end{equation}
In Eq.~\eqref{eq:fronteq}, $\tau = \dfrac{\sqrt{6} \xi c_s^2}{4E_0V_0^2}$ denotes the characteristic time scale of the front propagation, while $\beta > 0$ is a dimensionless constant arising from the derivation. The system is thus fully characterized by a single universal constant $\beta$ and a self-similar form $\Phi$, resulting from a nonlinear eigenvalue problem. Remarkably, the self-similar form is independent of the damping term.

Before further investigating Eqs.~\eqref{eq:fronteq} and \eqref{eq:Selfsimeq}, we test the self-similar form \eqref{eq:SelfSimAnsatz} using the measured values of $k_*(t)$. The compensated spectra collapse onto a single master curve (Fig.~\ref{fig:kstar}a) for all times. We present in the figure WKE simulations with and without damping, and GPE simulations with and without the confining trap. The excellent collapse confirms the existence of a universal self-similar regime.
Within this regime, the KZ scaling, anticipated earlier and confirmed in our numerical simulations, emerges as the unique asymptotic compatible with the similarity equation. Indeed, substituting a powerlaw ansatz $\Phi \sim \eta^{-x}$ ($x>0$) into Eq.\eqref{eq:Selfsimeq}, we see that the RHS vanishes much faster than the LHS as $\eta \to 0$ (see SM). We thus conclude that the spectrum tends to solutions of $\Phi(\eta) + \eta \Phi'(\eta) = 0$, i.e. $\Phi \sim \eta^{-1}$ (see SM). Interestingly enough, this corresponds to the KZ spectrum for which the RHS also vanishes. The scaling function $\Phi$ can be further characterized through the use of a differential approximation~\cite{hasselmann1985computations, nazarenko2006sandpile, nazarenko2006differential} of Eq.~\eqref{eq:Selfsimeq}, leading to a diffusive equation for large $\eta$:
\begin{equation*}
    -\beta \left[\Phi(\eta) + \eta \Phi'(\eta)\right] = D \eta^3\dfrac{d}{d\eta}\left(\eta^{-2}\Phi'(\eta)\right),
    \label{eq:Phieq}
\end{equation*}
with $D = 2\int_0^\epsilon \eta \Phi(\eta)d\eta$ the diffusion coefficient. The solution of the above equation with an UV-converging energy integral leads to an exponential asymptotic behavior at small scales
\begin{equation}
    \Phi(\eta) \underset{\eta \gg 1}{\sim} \exp\left[-\dfrac{\beta}{D}\eta\right],
    \label{eq:Asympt}
\end{equation}
in excellent agreement with our numerical simulations (Fig.~\ref{fig:kstar}a, inset).

While the collapse of $\Phi$ confirms the validity of the self-similar structure and fixes its functional form, the dynamics are governed by a single, universal, eigenvalue $\beta$ governing the front propagation. Solving Eq.~\eqref{eq:fronteq} yields a front propagating as
\begin{equation}
    k_*(t)=
    \begin{cases}
        k_0\exp{\left[\sqrt{\dfrac{2\beta t}{\tau}}\right]}   &\text{if } \gamma = 0 \\[6pt]
        k_0\exp{\left[\sqrt{\dfrac{2\beta\tau_D}{ \tau }(1-e^{-t/\tau_D})}\right ]}  &\text{else}
    \end{cases},
    \label{eq:kstar}
\end{equation}
which is in quantitative agreement with our numerical simulations of both GPE \& WKE (Fig.~\ref{fig:kstar}b), allowing for a direct determination of $\beta$. The numerical evolution of $k_*(t)$ is then fitted to Eq.~\eqref{eq:kstar}, leading to the universal value
    \begin{equation*}
        \beta \approx 8.9
    \end{equation*}
The extracted value of $\beta$ is found to be consistent across all simulations, thus confirming that the self-similar regime is fully characterized by the couple $(\Phi, \beta)$.

Although 2D acoustic waves represent one of the simplest wave turbulence system, their dynamics display rather unusual features. First, unlike other \textit{infinite capacity} systems, 2D acoustic waves do not exhibit the traditionally expected first kind self-similarity.
Indeed, dimensional analysis alone fails to determine the value of $\beta$, which instead emerges as an eigenvalue of the self-similar equation~\eqref{eq:Selfsimeq}. Second, the similarity equation~\eqref{eq:Selfsimeq} is independent of dissipation, implying universal dynamics. Such a surprising finding is a consequence of the fact that the degree of homogeneity of the dissipative term and the one of the collisional integral are equal. This universal feature may reflect the fact that 2D acoustic waves represent a borderline infinite capacity system, where the energy integral is only logarithmically divergent for the KZ scaling. The system thus sits at the crossover between first- and second-kind self-similarity, inheriting properties from both regimes.

In this Letter we have investigated the emergence of self-similar solutions in the two-dimensional GPE, that captures the dynamics of strong BECs (inviscid case) and polaritons (dissipative case). For strong condensate, and in the abscence of vortices, the GPE can be mapped into an acoustic wave turbulence problem described by the WKE. Using the WWTT, we have shown that the GPE admits self-similar solutions exhibiting characteristics of both first- and second-kind self-similarity, well described by the WWTT predictions. More strikingly, these self-similar solutions display universal behavior characterized by a dimensionless constant $\beta$, independent of the presence of dissipation. Using numerical simulations of the WKE, we measured this exponent from the front propagation and found $\beta \approx 8.9$. This establishes the present dynamics as an example of a self-similar solution of degenerate type, where the scaling exponents emerge from a combination of dimensional constraints and dynamical selection.

\begin{acknowledgments}
    This work was funded by the Simons Foundation Collaboration grant Wave Turbulence (Award ID 651471). This work was supported by the French government through the France 2030 investment plan managed by the National Research Agency (ANR), as part of the Initiative of Excellence Université Côte d’Azur under reference number ANR-15-IDEX-01. The authors are grateful to the Université Côte d’Azur’s Center for High-Performance Computing (OPAL infrastructure) for providing resources and support.
\end{acknowledgments}

\bibliography{apssamp}

\clearpage
\onecolumngrid
\clearpage
\begin{center}
{\large\bf Supplemental material : Universal self-similar evolution of two-dimensional acoustic turbulence in
Bose-Einstein condensates.}
\end{center}
\input{Supplementary}

\end{document}

%% file: Supplementary.tex
\section{Numerical setup}

\subsection{Gross-Pitaevskii simulations}

We perform numerical simulations of the GPE using the following confining trap
\begin{eqnarray}
    x_p&=&\frac{L}{\pi}\sin[(x-L/2)\pi/L]\\
    y_p&=&\frac{L}{\pi}\sin[(y-L/2)\pi/L]\\
    V_{\rm trap}(x, y)& =& V_0 \min\left(2 - \frac{1}{2}\left[1 - \tanh\left(\frac{x_p^2 - \left(\frac{L_x}{2}\right)^2}{(2\sigma)^2}\right)\right] - \frac{1}{2}\left[1 - \tanh\left(\frac{y_p^2 - \left(\frac{L_y}{2}\right)^2}{(2\sigma)^2}\right)\right], 1\right).
\end{eqnarray}
The functions $x_p$ and $y_p$ are approximations of the functions $x-L/2$ and $y-L/2$ around the center of the periodic box such that $x_p^2$ and $y_p^2$ are $L-$periodic. The parameters $L_x$ and $L_y$ set the size of trap. In simulations we set $V_0=5\mu$, $L_x=L_y=70\xi$, and $\sigma=2\xi$.
We then evolve the GPE using an imaginary time evolution to obtain an unperturbed ground state $\psi_{\rm trap}$

In order to perturb the system, we set in Fourier space a realization of random Bogoliubov waves in equipartition as follows
\begin{equation}
    b_{\bf k}=B e^{i\phi_k}/\sqrt{\omega_k}
\end{equation}
where $\phi_k$ are uniform independent random phases, $\omega_k$ is the Bogoliubov dispersion relation and  $B$ is a normalization constant fixing the amplitude of the perturbation.  We then use the Bogoliubov transformation to express the Fourier transform of $\psi$ in terms of $b_{\bf k}$. See, for instance, Eqs. S11 and S12 of the supplemental material of \cite{zhu2024turbulence} for an explicit expression in terms of $\xi$ and $c$. We then obtain the perturbation $\psi_{\rm pert}$ in the periodic box by performing an inverse Fourier transform. Finally, we use as initial condition for GP simulations in a trap $\psi_{\rm IC}=\psi_{\rm trap}\psi_{\rm pert}$.

\subsubsection{Energy spectrum}

In the main text we have defined the energy spectrum of the Gross-Pitaevskii runs as
\begin{equation}
    e_k=2c^2\xi^2\int_{|{\bf k}|=k}k^2 |\hat{\psi}_{\bf k}|^2\mathrm{d}{\bf k}.
\end{equation}
We notice that with this definition $\int_0^\infty e_k\mathrm{d}k=\frac{2 c^2 \xi^2}{L^2}\int |\nabla \psi|^2\mathrm{d}{\bf r}=2E_{\rm kin}$, i.e. twice the mean kinetic energy of the flow. Under the assumption of weak nonlinearity, internal and kinetic energy are in equipartition \cite{griffin2022energy}, so that the former definition coincides with the energy spectrum of the WWTT, i.e. it follows that $2E_{\rm kin}=E_0$. 

\subsection{Wave kinetic equation simulations}

        \label{sec:Num}

        \medbreak \textit{WKE - WavKinS:} We perform numerical simulations of the wave kinetic equation using WavKinS~\cite{krstulovic2025wavkins}. The WKE is solved on a logarithmic grid, with wave numbers following a geometric progression 
        \begin{equation*}
            k_n = k_{\min}\lambda^n.
        \end{equation*}
        The grid parameter $\lambda$ is set by the resolution
        \begin{equation*}
            \lambda(N, k_{\min}, k_{\max}) = (k_{\max} / k_{\min})^{1/N},
        \end{equation*}
        where $N$ is the number of grid points. Such progression allows for numerical simulations with wave numbers spanning several decades.

    \section{Self-similar equation}
        The dynamics of the energy spectrum are described by the 2D WKE
        \begin{equation}
            \begin{aligned}
                \partial_t e_k &=  \dfrac{4V_0^2}{\sqrt{6} c_s^2\xi}S_t[e_k] - \gamma e_k,\\
                S_t[e_k] &= \int_0^\infty \left(k_1k_2\right)^{-1} \left(\mathcal{R}_{1, 2}^k - 2\mathcal{R}_{k, 2}^1\right) \, dk_1, \\
                \mathcal{R}_ {1,2}^k &= H(k_2)\left(k^2e_1 e_2 - k_1^2e_2e_k - k_2^2 e_1e_k\right),
            \end{aligned}
            \label{eq:SimpleWKEsup}
        \end{equation}
        where $k_2 = |k-k_1|$, $V_0 = \dfrac{3}{4}\sqrt{\dfrac{c_s}{2}}$ and $H$ is the Heaviside function.
        \medbreak\noindent We look for self-similar solutions of the form
        \begin{align*}
            e_k(t) \equiv E_0f(t) \Phi(\eta)&& \eta = k/k_*.
        \end{align*}
        Substituting the above definition into Eq.~\eqref{eq:SimpleWKEsup} and using the
        energy definition yields
        \begin{align}
         &\left(\frac{f'}{f}+\gamma\right)\Phi -\frac{k_*'}{k_*}\,\eta\,\Phi' =\frac{f k_*}{\tau}\,S_t[\Phi], \\
        &E = E_0 e^{-t/\tau_D}
           \simeq E_0 f(t) k_*(t)\ln\!\left(\frac{k_*}{k_0}\right),
           \qquad (k_* \gg k_0), \\
        &\tau = \dfrac{\sqrt{6} \xi c_s^2}{4E_0V_0^2}, \qquad\tau_D = \gamma^{-1}. \notag
        \end{align}
       The logarithmic factor $\ln(k_*/k_0)$ arises from the asymptotic behavior $\Phi(\eta)\sim \eta^{-1}$ as $\eta\to 0$, which dominates the energy integral.
        The energy integral then gives
        \begin{equation*}
            f(t)=\frac{e^{-t/\tau_D}}{k_*\,\ln(k_*/k_0)} .
        \end{equation*}
        Plugging the above result into the self-similar equation leads to
        \begin{equation}
           - \left[\left( \dfrac{k_*'}{k_*}\ln\left(\dfrac{k_*}{k_0} \right) + \dfrac{1}{k_*}\right)\Phi + \dfrac{k_*'}{k_*}\ln\left(\dfrac{k_*}{k_0} \right)\eta\Phi'\right] = \dfrac{e^{-t/\tau_D}}{\tau}S_t\left[\Phi\right].
        \end{equation}
        The above equation is only valid in the limit $k_* \gg k_0$ such that it simplifies to 
        \begin{equation}
           - \dfrac{k_*'
            }{k_*}\ln\left(\dfrac{k_*}{k_0} \right)\left[\Phi + \eta\Phi'\right] = \dfrac{\exp\left[-t/\tau_D\right]}{\tau}S_t\left[\Phi\right].
        \end{equation}
        The propagation of the front is thus obtained by canceling the time dependence leading to
        \begin{align}
            - \beta\left[\Phi(\eta) + \eta\Phi'(\eta)\right] &= S_t\left[\Phi\right], \\
            \dfrac{k_*'}{k_*}\ln\left(\dfrac{k_*}{k_0} \right) &= \dfrac{\beta}{\tau}\exp\left[-t/\tau_D\right].
        \end{align}

    \section{Front definition}
        We define the front as 
        \begin{equation}
            k_*(t) = \left({\int_0^\infty k^{n+2}e_k\mathrm{d}k}\Big/{\int_0^\infty k^n e_k\mathrm{d}k} \right)^{1/2}.
        \end{equation}
        While a natural choice would be $n=0$, this leads to an IR-divergent denominator since $e_k \underset{k\ll1}{\sim} k^{-1}$. We therefore choose the smallest value of $n$ that removes this divergence, yielding
        \begin{equation}
            k_*(t) = \left({\int_0^\infty k^{3}e_k\mathrm{d}k}\Big/{\int_0^\infty k e_k\mathrm{d}k} \right)^{1/2}.
        \end{equation}
    \section{Differential approximation}
        The energy spectrum $e_k(t)$ satisfies the integral equation
        \begin{equation*}
            \begin{aligned}
                \left( \dfrac{4V_0^2}{\sqrt{6} c_s^2\xi}\right)^{-1}\partial_te_k &= \int_0^k \left(k_1k_2\right)^{-1} \mathcal{R}_{1,2}^k \, dk_1 - 2\int_k^\infty \left(k_1k_2\right)^{-1}\mathcal{R}_{k,2}^1 \, dk_1, \\
                \mathcal{R}_{1,2}^k &= \left(k^2e_1e_2 - k_1^2 e_ke_2 - k_2^2 e_ke_1\right),
            \end{aligned}
        \end{equation*}
        where $k_2 = |k-k_1|$. While solving such equation is rather non trivial, one can obtain the asymptotical behavior of $e_k$ at large $k$ using the differential approximation. Let $k \rightarrow \infty$, the first integral $\int_0^k \left(k_1k_2\right)^{-1} \mathcal{R}_{1,2}^k \, dk_1$ can be rewritten using the $1 \leftrightarrow 2$ symmetry such that either $k_2 \rightarrow\infty$ or $k_1 \rightarrow \infty$ yielding:
        \begin{equation*}
            \int_0^k \left(k_1k_2\right)^{-1} \mathcal{R}_{1,2}^k \, dk_1 \approx 2\int_0^\epsilon \left(k_1k_2\right)^{-1} \mathcal{R}_{1,2}^k \, dk_1.
        \end{equation*}
        The second integral only allows for large values of $k_1$ such that in the large $k$ limit, the collision integral reads:
        \begin{equation}
            S_t\left[e\right] = 2\int_0^\epsilon \left[(k - k_1)k_1\right]^{-1} \mathcal{R}_{1,2}^k \, dk_1 - 2\int_0^\epsilon \left[(k + k_2)k_2\right]^{-1}\mathcal{R}_{k,2}^1 \, dk_2,
            \label{eq:coll}
        \end{equation}
        Using that $\underset{k \rightarrow\infty}{\lim}e_k \ll1$, one can simplify the integrands:
        \begin{equation*}
            \begin{cases}
                \mathcal{R}_{k,2}^1 \approx e_1\left(k^2 e_{k-k_1} - (k-k_1)^2e_k\right) ,\\
                \mathcal{R}_{1,k}^2 \approx e_2\left((k+k_2)^2 e_{k} - k^2e_{k+k_2}\right)
            \end{cases}
        \end{equation*}
        Performing Taylor-expansions and substituting in Eq.~\ref{eq:coll} yields the diffusion equation:
        \begin{equation}
            \partial_te_k = D k^3\dfrac{\partial}{\partial k}\left(k^{-2}\dfrac{\partial e_k}{\partial_k}\right),
            \label{eq:Phieqsup}
        \end{equation}
        where $D = \dfrac{8V_0^2}{\sqrt{6} c_s^2\xi}\int_0^\epsilon k e_k dk$ is the diffusion coefficient. Substituting the self-similar form into Eq.~\eqref{eq:Phieqsup} and considering the following ansatz: 
        \begin{align*}
            &\Phi \underset{\eta \gg 1}{\sim} \eta^b, \\
            &\Phi \underset{\eta \gg 1}{\sim} \exp\left[\lambda \eta^b\right], \qquad \lambda \in \mathbb{R}
        \end{align*}
        yields, at leading order
        \begin{equation*}
            \begin{cases}
                \beta(b+1)\eta^b = 0, \\[12pt] 
            \lambda Db \eta^{2b-1} + \beta \eta^b = 0.
            \end{cases}  \Leftrightarrow \quad
            \begin{cases}
                &\Phi \underset{\eta \gg 1}{\sim} \eta^{-1}, \\
            &\Phi \underset{\eta \gg 1}{\sim} \exp\left[-\dfrac{\beta}{D}\eta\right].
            \end{cases}
        \end{equation*}
        In addition, one must have a UV-converging energy integral $\int \Phi d\eta$ such that the powerlaw solution has to be discarded. Finally, one has:
        \begin{equation}
            \Phi \underset{\eta \gg 1}{\sim} \exp\left[-\dfrac{\beta}{D}k\right].
        \end{equation}